\documentclass [12pt] {article}
\usepackage{a4}
 
\usepackage{amsfonts}
\usepackage{amsmath} 
\usepackage{graphics}
\usepackage{epsfig}
\usepackage{epic}

\author{T.~Leonhardt and W.~R\"uhl}
\title{The minimal conformal O(N) vector sigma model at d=3}

\begin{document}

\newcommand {\eps}{\varepsilon}
\newcommand {\ti}{\tilde}
\newcommand {\D}{\Delta}
\newcommand {\G}{\Gamma}
\newcommand {\de}{\delta}
\newcommand {\al}{\alpha}
\newcommand {\la}{\lambda}
\newcommand {\ind}{\mathrm{d}}
\thispagestyle{empty}

\noindent hep-th/0308111   \hfill August 2003 \\                   
 
\noindent
\vskip3.3cm
\begin{center}
 
{\Large\bf The minimal conformal $\boldsymbol{O(N)}$ vector sigma model at $\boldsymbol{d=3}$}
\bigskip\bigskip\bigskip
 
{\large Thorsten Leonhardt and Werner R\"uhl}
\medskip
 
{\small\it Department of Physics\\
     Erwin Schr\"odinger Stra\ss e \\
     University of Kaiserslautern, Postfach 3049}\\
{\small\it 67653 Kaiserslautern, Germany}\\
\medskip
{\small\tt tleon,ruehl@physik.uni-kl.de}
\end{center}
 
\bigskip 
\begin{center}
{\sc Abstract}
\end{center}
\noindent For the minimal $O(N)$ sigma model, which is defined to be generated by the $O(N)$ scalar auxiliary field alone, all $n$-point functions, till order $1/N$ included, can be expressed by elementary functions without logarithms. Consequently, the conformal composite fields of $m$ auxiliary fields possess at the same order such dimensions, which are $m$ times the dimension of the auxiliary field plus the order of differentiation.

\newpage

\section{Introduction}

The conformal $O(N)$ vector sigma model has been studied in a series of basic works in \cite{Wilson:1973jj,Vasilev:yc,Vasilev:dg,Vasilev:dc} and then been analyzed from several points of view in \cite{Lang:1991kp,LR,Lang:1992pp}. Recently it has attracted interest as a candidate for an AdS$_4$/CFT$_3$ correspondence \cite{Klebanov:2002ja}, where the AdS$_4$ theory is a special higher spin gauge field theory on AdS space of the type investigated in a large series of papers by M.~Vasiliev \cite{Fradkin:ks,Fradkin:1986qy,Vasiliev:1995dn,Vasiliev:2003ev}, see also \cite{Fronsdal:1978rb,Sezgin:2002ru}.

We want to describe here the properties of the ``minimal'' interacting model, consisting only of the ``auxiliary'' or ``Lagrange multiplier'' field $\al$, which is a scalar under Lorentz and internal $O(N)$ transformations. Thus we are interested in the set of $n$-point functions $\langle \al(x_1) \cdots \al(x_n) \rangle$, where we consider the ``physical'' spacetime dimension $d=3$, because in the references given above, $d$ is a parameter in the open interval $d\in (2,4)$. The $n$-point functions are given by a $1/N$-expansion. We shall describe the simplifications arising by the restrictions to $d=3$. They allow us to describe any $n$-point function of $\al$ fields up to (and including) the perturbative order $O(1/N)$ in explicit and simple form.

Besides the interacting conformal model a free theory based solely on the free $O(N)$ vector field $\phi = (\phi_1,\ldots,\phi_n)$, which is also a scalar under Lorentz transformations, enables us to calculate the $n$-point functions of the ``scalar current''
\begin{align}
J(x) = \sum_{j=0}^N \phi_j(x) \phi_j(x).
\end{align}
If we perturb the free $O(N)$ model by the ``double-trace'' operator
\begin{align}
\frac{\la}{2 N} J(x)^2,
\end{align}
the free theory flows from the corresponding unstable ultraviolet fixed point to the stable infrared fixed point \cite{Witten:2001ua,Gubser:2002vv}, which belongs to the interacting $O(N)$ theory mentioned above. During this flow, the field dimension of $J$ changes from its free field value $\D_- = 1$ to the field dimension $\D_+=2$ of $\al$. Both fixed points are connected by the AdS/CFT correspondence to the same bulk field with mass $m^2=-2$,  therefore we have
\begin{align}
\D_\pm = \mu \pm \sqrt{\mu^2-2},
\end{align}
where we introduced the convenient abbreviation $\mu= d/2$, which will be used throughout the article. The theories at both fixed points are connected by a Legendre transformation\cite{Klebanov:1999tb}, see section \ref{legendre}.


\section{The $\boldsymbol{O(N)}$ model at its interacting fixed point}

A perturbative expansion of a CFT is formulated by a skeleton graph expansion, where the propagators and the three-point vertices are essentially fixed by conformal covariance. In the $O(N)$ sigma model at the infrared stable fixed point we thus have the full propagators for the auxiliary field $\al$ and the $O(N)$ vector $\phi$
\begin{align}\label{props}
G(x_{12}) & := \langle \alpha(x_1) \alpha(x_2) \rangle = \bigl(x_{12}^2 \bigr)^{-\beta}, \;\textrm{denoted by a dashed line}, \nonumber \\
D(x_{12}) & := \langle \phi_i(x_1) \phi_j(x_2) \rangle = \delta_{ij}\bigl( x_{12}^2
\bigr)^{-\de}, \;\textrm{denoted by a solid line}.
\end{align}
The latter must be taken into account in the internal lines of the skeleton graphs for the $n$-point functions of $\al$. We mention that the normalization constants of the two-point functions are absorbed in the coupling constant $z$ of the interaction vertex, which is given by 
\begin{align}\label{intvert}
z^{\frac{1}{2}} \int \ind x \;\al(x) \sum_{j=1}^N \phi_j(x)^2.
\end{align}
$z$ assumes a fixed value at the interacting fixed point (critical value), which has the expansion
\begin{align}
z= \sum_{i=1}^\infty \frac{z_i}{N^i}.
\end{align}

The field dimensions $\beta$ and $\de$ decompose into a canonical and an anomalous part $\eta$,
\begin{align}
\beta &= 2 + \eta(\al) \\
\de &= \mu-1+\eta(\phi),
\end{align}
where the anomalous parts can be expanded in $1/N$:
\begin{align}
\eta(\al) & = \sum_{i=1}^\infty \frac{\eta_i(\al)}{N^i} \\
\eta(\phi) & = \sum_{i=1}^\infty \frac{\eta_i(\phi)}{N^i}.
\end{align}
The expansion coefficients $\eta_1(\phi), \eta_2(\phi), \eta_3(\phi)$ and $\eta_1(\al)$ are known \cite{Vasilev:yc,Vasilev:dg,Vasilev:dc,Bernreuther:js}. 

In general, an integral
\begin{align}
\int \ind y \; \prod_{i=1}^M \Bigl( (y-x_i)^2 \Bigr)^{-\gamma_i}
\end{align}
is conformally invariant if the dimensions $\gamma_i$ satisfy the constraint of ``uniqueness''
\begin{align}
\sum_{i=1}^M \gamma_i = d.
\end{align}
However, the interaction vertex (\ref{intvert}) has to be UV renormalized, thereby violating uniqueness, i.e. $\beta + 2 \de =d-2 \kappa$ implies $\eta(\al)= -2 \eta(\phi) -2 \kappa$, where the deviation from uniqueness $\kappa$ can be expanded in a series in $1/N$, similarly to $\eta$.

Let us explain how the coupling constants $z$ and the anomalous dimension $\eta(\phi)$ can be obtained from the conformal bootstrap equations
\begin{align}\label{bootstrap}
D^{-1} & + \begin{picture}(42,15)
\put(4,2){
\put(0,0){\dashline{4}(0,0)(33,0)}
\circle{3}
\put(30,0){\circle{3}}
\put(-3,0){\qbezier(0,0)(16,12)(33,0)}}
\end{picture}
+
\begin{picture}(42,20)
\put(4,3){
\put(0,0){\dashline{4}(0,0)(16,-16)}
\put(33,0){\dashline{4}(0,0)(-16,16)}
\circle{3}
\put(30,0){\circle{3}}
\put(30,0){\line(-1,-1){17}}
\put(13.5,16){\line(0,-1){32}}
\put(-3,0){\line(1,1){16}}}
\end{picture}
+ \cdots = 0 
\end{align}

\vspace{-.2cm}

\begin{align}\label{boot2}
\frac{2}{N} \, G^{-1} & + \begin{picture}(42,15)
\put(4,2){
\put(0,0){\qbezier(0,0)(16,-12)(33,0)}
\circle{3}
\put(30,0){\circle{3}}
\put(-3,0){\qbezier(0,0)(16,12)(33,0)}}
\end{picture}
+
\begin{picture}(42,20)
\put(4,3){
\put(16.3,16){\dashline{4}(0,0)(0,-32)}
\circle{3}
\put(29.2,0){\circle{3}}
\put(29,0){\line(-1,1){16}}
\put(29,0){\line(-1,-1){16}}
\put(-3,0){\line(1,-1){16}}
\put(-3,0){\line(1,1){16}}}
\end{picture}
+ \cdots = 0, 
\end{align}
which are equations for the respective amputated two-point functions. The amputations are performed by the inverses of the propagators (\ref{props}). Explicitly, the propagators are kernels of the form
\begin{align}
F(x) =(x^2)^{-\la}
\end{align}
and the respective inverses $F^{-1}$ are defined by 
\begin{align}
\int \ind x_2 F(x_{12})F^{-1}(x_{23}) = \de(x_{13}).
\end{align}
Then one easily gets
\begin{align}\label{normal}
F^{-1} (x) = p(\la) (x^2)^{-(d-\la)}
\end{align}
with 
\begin{align}
p(\la) = \pi^{-2\mu} \frac{a(\la-\mu)}{a(\la)}; \quad a(\la)= \frac{\G(\mu-\la)}{\G(\la)}.
\end{align}

Now we insert (\ref{props}) and (\ref{intvert}) into the bootstrap equations (\ref{bootstrap},\ref{boot2}), restrict to the leading terms, respectively, and obtain
\begin{align}\label{bootatfirst}
p(\de) & = - z \nonumber \\
p(\beta) &= -\frac{1}{2} N z .
\end{align}
The second equation in (\ref{bootatfirst}) implies 
\begin{align}
z_1 = -2 p(2) = 2 \pi^{-2\mu} \frac{(\mu-2) \G(2\mu-2)}{\G(\mu) \G(1-\mu)},
\end{align}
whereas the first equation in (\ref{bootatfirst}) gives
\begin{align}
\eta_1(\phi) = 2 \frac{\sin \pi \mu}{\pi} \frac{\G(2\mu-2)}{\G(\mu+1) \G(\mu-2)}.
\end{align}
At $d=3$ we furthermore obtain
\begin{align}
\eta_1(\phi)\Bigl \rvert_{d=3} = \frac{4}{3}, \quad \eta_1(\al)\Bigl \rvert_{d=3} = - \frac{32}{3}, \quad \kappa_1\Bigl \rvert_{d=3} = 4.
\end{align}

\section{Free field theory and a Legendre transform}\label{legendre}

We consider the free field theory of the composite fields (normal ordering understood)
\begin{align}
J(x) =  \sum_{i=1}^N \phi_i(x) \phi_i(x).
\end{align}
With the normalization of the $\phi$ field (\ref{props}) we obtain for the two-, three- and four-point function of $J$ by Wick's theorem
\begin{align}
\langle J(x_1) J(x_2) \rangle &= 2 N (x_{12}^2)^{-d+2} \\
\langle J(x_1) J(x_2) J(x_3) \rangle &= 8N (x_{12}^2 x_{23}^2 x_{13}^2)^{-(\mu-1)} \label{threeptJ}
\end{align}
and
\begin{multline}
\langle J(x_1) J(x_2) J(x_3) J(x_4) \rangle = 4N^2 \Bigl[ (x_{12}^2 x_{34}^2)^{-d+2} + (x_{13}^2 x_{24}^2)^{-d+2} + (x_{14}^2 x_{23}^2)^{-d+2} \Bigr] \\
+ 16 N \Bigl[(x_{12}^2 x_{23}^2 x_{34}^2 x_{41}^2)^{-\mu+1} + (x_{12}^2 x_{24}^2 x_{43}^2 x_{31}^2)^{-\mu+1} + (x_{13}^2 x_{32}^2 x_{24}^2 x_{41}^2)^{-\mu+1} \Bigr].
\end{multline}  
From \cite{Klebanov:2002ja} we have learned that this theory is connected to the interacting sigma model by a Legendre transformation with respect to the dual field $\al$ \footnote{In contrast to the previous section, we now choose to set the coupling constant of the three vertex to unity, which implies that the normalization of the two-point function is the one given in (\ref{normal})},
\begin{align}
\int \ind x J(x) \al(x).
\end{align}
Thus let us perform this Legendre transformation. To this end, we build up every diagram of $\al$ fields that can be constructed from the $n$-point functions of $J$ by attaching $\al$ propagators, normalized as 
\begin{align}
\langle \al(x_1) \al(x_2) \rangle = K (x_{12})^{-2},
\end{align}
to the respective legs. Since $\al$ is dual to $J$, the respective two-point functions are inverses to each other
\begin{align}
\int \ind x_3 (x_{13}^2)^{-d+2} (x_{32}^2)^{-2} = p(2)^{-1} \de(x_{12}).
\end{align}
Consequently, we get the $\al$ two-point function from the Legendre transformed $J$ two-point function
\begin{align}
\int \ind x_3 \ind x_4 \langle \al(x_1) \al(x_3) \rangle \langle J(x_3) J(x_4) \rangle \langle \al(x_4) \al(x_2) \rangle = 2 N K^2 p(2)^{-1} (x_{12}^2)^{-2},
\end{align}
if we use the normalization constant (see (\ref{normal}))
\begin{align}
K=\frac{1}{2} \Bigl( \frac{z_1}{N} \Bigr)^{\frac{1}{2}}.
\end{align}

Performing the Legendre transformation on the three-point function of $J$ (\ref{threeptJ}) gives the $\al$ three-point function
\begin{align}
\langle \al (x_1) \al (x_2) \al(x_3) \rangle = \begin{picture}(60,30)
\put(8,10){
\dashline{4}(20,10)(20,-10)
\put(20,-10){\line(-1,-1){10}}
\put(20,-10){\line(1,-1){10}}
\put(10,-20){\line(1,0){20}}
\dashline{4}(10,-20)(0,-30)
\dashline{4}(30,-20)(40,-30)
\put(-10,-35){$2$}
\put(18,15){$1$}
\put(45,-35){$3$}}
\end{picture}\quad ,
\end{align}
\vspace{.1cm}

\noindent which can be integrated to give 
\begin{align}\label{althreept}
\langle \al (x_1) \al (x_2) \al(x_3) \rangle = N \Bigl( \frac{z_1}{N} \Bigr)^{\frac{3}{2}} v(2, \mu-1, \mu-1)^2 v(2,1,2\mu-3) (x_{12}^2 x_{13}^2 x_{23}^2)^{-1}, 
\end{align}
where 
\begin{align}
v(\al_1,\al_2, \al_3) = \pi^\mu \prod_{i=1}^3 \frac{\G(\mu-\al_i)}{\G(\al_i)}.
\end{align}
Now we observe that $v(2,1,2 \mu-3)$ in (\ref{althreept}) contains a factor $\G(2\mu-3)$ in the denominator, thus the $\al$ three-point function vanishes at $d=3$. 

The Legendre transform of the $J$ four-point function gives the $\al$ four-point function, which consists of three disconnected graphs
\begin{align}\label{disconngraph}
\begin{picture}(60,40)
\put(20,-20){
\dashline{4}(-30,50)(20,50)
\dashline{4}(-30,0)(20,0)
\put(-40,46){$1$}
\put(25,46){$2$}
\put(-40,-4){$3$}
\put(25,-4){$4$}
\put(-10,-20){$A_1$}}
\end{picture}
\qquad\qquad
\begin{picture}(60,40)
\put(20,-20){
\dashline{4}(-30,50)(-30,0)
\dashline{4}(20,50)(20,0)
\put(-40,46){$1$}
\put(25,46){$2$}
\put(-40,-4){$3$}
\put(25,-4){$4$}
\put(-10,-20){$A_2$}}
\end{picture}
\qquad\qquad
\begin{picture}(60,40)
\put(20,-20){
\dashline{4}(-30,50)(20,0)
\dashline{4}(-30,0)(-11,19)
\dashline{4}(-1,29)(20,50)
\put(-40,46){$1$}
\put(25,46){$2$}
\put(-40,-4){$3$}
\put(25,-4){$4$}
\put(-10,-20){$A_3$}}
\end{picture},
\end{align}
\vspace{8mm}

\noindent three box graphs

\begin{align} \label{boxgraphsforO(N)}
\begin{picture}(80,30)
\put(10,10){
\dashline{4}(0,30)(10,20)
\dashline{4}(50,30)(40,20)
\put(10,20){\line(1,0){30}}
\put(10,20){\line(0,-1){30}}
\dashline{4}(0,-20)(10,-10)
\dashline{4}(50,-20)(40,-10)
\put(40,-10){\line(-1,0){30}}
\put(40,-10){\line(0,1){30}}
\put(-8,-25){$3$}
\put(-8,28){$1$}
\put(53,-25){$4$}
\put(53,28){$2$}
\put(17,-40){$B_{21}$}}
\end{picture}
\qquad  \quad
\begin{picture}(80,30)
\put(5,10){
\dashline{4}(0,30)(10,20)
\dashline{4}(60,30)(40,-10)
\put(10,20){\line(1,0){30}}
\put(10,20){\line(0,-1){30}}
\dashline{4}(0,-20)(10,-10)
\dashline{4}(60,-20)(50,-1)
\dashline{4}(45,9)(40,20)
\put(40,-10){\line(-1,0){30}}
\put(40,-10){\line(0,1){30}}
\put(-8,-25){$3$}
\put(-8,28){$1$}
\put(63,-25){$4$}
\put(63,28){$2$}
\put(17,-40){$B_{22}$}}
\end{picture}
\qquad \quad
\begin{picture}(80,30)
\put(-5,15){
\dashline{4}(0,30)(10,20)
\dashline{4}(50,30)(40,20)
\put(10,20){\line(1,0){30}}
\put(10,20){\line(0,-1){30}}
\dashline{4}(0,-30)(40,-10)
\dashline{4}(20,-15)(10,-10)
\dashline{4}(50,-30)(32,-21)
\put(40,-10){\line(-1,0){30}}
\put(40,-10){\line(0,1){30}}
\put(-8,-35){$3$}
\put(-8,28){$1$}
\put(53,-35){$4$}
\put(53,28){$2$}
\put(17,-45){$B_{23}$}}
\end{picture},
\end{align}
\vspace{5mm}

\noindent which will be treated in detail in the next section and the one-particle reducible graphs
\vspace{5mm}
\begin{align} \label{exchgraphsforO(N)}
\begin{picture}(60,30)
\put(-10,10){
\dashline{4}(0,30)(10,20)
\dashline{4}(30,20)(40,30)
\put(10,20){\line(1,0){20}}
\put(10,20){\line(1,-1){10}}
\put(30,20){\line(-1,-1){10}}
\dashline{4}(20,10)(20,-10)
\put(20,-10){\line(-1,-1){10}}
\put(20,-10){\line(1,-1){10}}
\put(10,-20){\line(1,0){20}}
\dashline{4}(10,-20)(0,-30)
\dashline{4}(30,-20)(40,-30)
\put(-10,-35){$3$}
\put(-10,28){$1$}
\put(45,-35){$4$}
\put(45,28){$2$}
\put(15,-40){$B_{11}$}}
\end{picture}
\qquad \qquad\!\!\!
\begin{picture}(60,30)
\put(-0,0){
\dashline{4}(-10,30)(0,20)
\dashline{4}(-10,-10)(0,0)
\line(0,1){20}
\put(0,0){\line(1,1){10}}
\put(0,20){\line(1,-1){10}}
\dashline{4}(10,10)(30,10)
\put(30,10){\line(1,1){10}}
\put(30,10){\line(1,-1){10}}
\put(40,0){\line(0,1){20}}
\dashline{4}(40,0)(50,-10)
\dashline{4}(40,20)(50,30)
\put(-19,-15){$3$}
\put(-19,28){$1$}
\put(53,-15){$4$}
\put(53,28){$2$}
\put(15,-30){$B_{12}$}}
\end{picture}
\qquad\qquad
\begin{picture}(60,30)
\put(-0,10){
\dashline{4}(-10,30)(10,-20)
\dashline{4}(30,20)(40,30)
\put(10,20){\line(1,0){20}}
\put(10,20){\line(1,-1){10}}
\put(30,20){\line(-1,-1){10}}
\dashline{4}(20,10)(20,-10)
\put(20,-10){\line(-1,-1){10}}
\put(20,-10){\line(1,-1){10}}
\put(10,-20){\line(1,0){20}}
\dashline{4}(10,20)(5,8)
\dashline{4}(0,-5)(-10,-30)
\dashline{4}(30,-20)(40,-30)
\put(-20,-35){$3$}
\put(-20,28){$1$}
\put(45,-35){$4$}
\put(45,28){$2$}
\put(10,-40){$B_{13}$}}
\end{picture},
\end{align}
\vspace{5mm}

\noindent which have been evaluated in \cite{Lang:1992pp}
 and are quoted there in the appendix. At $d=3$, these graphs have a second order zero, which seem to originate from each $\al$ three-point function contained in the graph.


\section{The four-point function in the interacting conformal sigma model and its $\boldsymbol{d=3}$ limit}

Now we consider the sigma model at its interacting critical point. This means we have to calculate the four-point function $\langle \al(x_1) \al(x_2) \al(x_3) \al(x_4) \rangle$ up to order $1/N$, which consists of the three types of graphs mentioned in the previous section: the disconnected graphs, the 1-P-reducible graphs, which vanish at $d=3$, and the boxgraphs, which are investigated in this section.
In \cite{Lang:1992pp}
, they are only partly computed  and, as far as we know,  have never been fully evaluated. Let us repeat the results of this reference: The boxgraphs $B_{2j},\; j=1,2,3,$ have the structure
\begin{align}
B_{2j} = (x_{12}^2 x_{34}^2)^{-2} \sum_{m,n \ge0} \frac{u^n (1-v)^m}{n!\;m!} \Bigl[ -a_{nm}^{(2j)} \log u + b_{nm}^{(2j)} + u^{\mu-3} c_{nm}^{(2j)} \Bigr], 
\end{align}
see eqns. (5.11)-(5.13) of \cite{Lang:1992pp}, where the coefficients $a^{(2j)}_{nm}$ are given in eqns. (C.7), (C.9) and (C.11) in \cite{Lang:1992pp}. The conformally invariant variables $u,v$ are defined by 
\begin{align}
u=\frac{x_{13}^2 x_{24}^2}{x_{12}^2 x_{34}^2}; \quad v=\frac{x_{14}^2 x_{23}^2}{x_{12}^2 x_{34}^2}.
\end{align}
We observe, that all of them contain a factor $\G(2\mu-4)^{-1}$, giving a zero at $d=3$. The coefficients $c_{nm}^{(2j)},\; j=1,2$ (eqns. (C.8) and (C.10)\footnote{In this equation there is a wrong factor of $1/n!$, which should be removed.}) have even a double zero from the $\G(2\mu-3)^{-2}$, but this zero is compensated by a double pole for all but a few $(n,m)$. These surviving terms lead to current exchanges, as discussed in \cite{Leonhardt:2002sn}. We shall return to them below. Moreover, we also have $c_{nm}^{(23)}=0$.

The coefficients $b_{nm}^{(2j)},\; j=1,2,3,$ are not completely evaluated, but they all have a zero at $d=3$. To see this, let us turn to the general boxgraph (\ref{genbox}), which has been discussed in a more general framework in \cite{Lang:1991kp}. 

\newpage

\begin{align} \label{genbox}
\begin{picture}(80,50)
\put(0,0){
\put(0,0){\line(1,0){50}}
\put(0,0){\line(0,1){50}}
\put(50,0){\line(0,1){50}}
\put(0,50){\line(1,0){50}}
\put(-15,22){$\beta_4$}
\put(53,22){$\beta_2$}
\put(20,-12){$\beta_3$}
\put(20,55){$\beta_1$}
\dashline{4}(0,0)(-30,-30)
\dashline{4}(0,50)(-30,80)
\dashline{4}(50,50)(80,80)
\dashline{4}(50,0)(80,-30)
\put(-29,-12){$\al_3$}
\put(-29,57){$\al_1$}
\put(68,-12){$\al_4$}
\put(68,57){$\al_2$}
\put(-37,-35){$3$}
\put(-37,79){$1$}
\put(82,-35){$4$}
\put(82,79){$2$}
}
\end{picture}
\end{align}

\vspace{1cm}

\noindent The integration technique consisted in doing two unique three-vertex integrals at opposite positions in the graph, performing one of the two remaining four-vertex integrals and transforming the second one into a Barnes type integral via Symanzik's technique \cite{Symanzik:1972wj}. This results in 
\begin{align}
(x_{13}^2)^{-\al_1} (x_{34}^2)^{\al_1-\al_3} (x_{24}^2)^{-\al_2} B(u,v)
\end{align}
with
\begin{align}\label{FormelB}
B(u,v)= \sum_{m,n=0}^\infty \frac{u^n (1-v)^m}{n!\;m!} \Bigl[ u^{\al_1} c_{nm}^{(1)} + u^{\al_2} c_{nm}^{(2)} + u^{\beta_3} c_{nm}^{(3)} \Bigr].
\end{align}
The non-evaluated part of this graph is a contribution to $c_{nm}^{(2)}$ (see eq. (A.9) in \cite{Lang:1991kp})
\begin{multline}\label{crucint}
c_{nm,1}^{(2)} = \pi^{2\mu} \frac{v(\al_1, \beta_1, \beta_4) v( \al_4, \beta_2, \beta_3)}{\prod_{j=1}^4 \G(\delta_j)} (2 \pi i)^{-2} \int_{-i \infty}^{i \infty} \ind x \int_{-i \infty}^{i \infty} \ind y \, \G(-x) \G(-y) \\ \G(\de_4+x+y) \G(\mu -\de_1+x+y) 
\G(\de_1 + \de_2 -\mu-y) \G(\de_1 + \de_3- \mu-x) \\
a(\gamma_1) a(\gamma_3) a(\gamma_2+\gamma_4) \frac{(\gamma_4)_n (\mu-\gamma_3)_n (\gamma_2)_{n+m} (\mu-\gamma_1)_{n+m}}{(\gamma_2 + \gamma_4)_{2n+m} (\gamma_2+\gamma_4 -\mu +1)_n},
\end{multline}
where 
\begin{align}
\de_1 & = \mu- \beta_1, & \de_2 &= 2\mu-\al_1-\al_2, & \de_3 &= \al_3, & \de_4 &= \mu-\beta_2, \\
\gamma_1 &= \mu -\al_4-y, & \gamma_2 &= \al_2, & \gamma_3 &= \beta_1+x+y, & \gamma_4 &= \mu -\beta_3 - x.
\end{align}
The crucial factor is $\G(\de_2)^{-1}$, since at the end we set $\al_1=\al_2=2$. This produces a simple zero at $d=3$. The question is whether this zero is canceled by a pole. 

If 
\begin{align}
\D:=\al_1-\al_2
\end{align}
tends to zero, then poles in $\D$ in $c_{nm}^{(1)}$ and $c_{nm}^{(2)}$ (\ref{FormelB}) arise, which cancel each other and give
\begin{align}
u^{\al_1} \Bigl[ -a_{nm} \log u + b_{nm} \Bigr].
\end{align}
These poles are obviously independent of the pole at $2 \mu -\al_1-\al_2 = 2 \mu-4 =-1$ and can be neglected in this context. The standard way of evaluating the integral (\ref{crucint}) is by shifting the contours to $+ \infty$ and summing up the residues. This results in generalized hypergeometric series of argument one, whose poles in the parameters are difficult to evaluate.

Thus we used the method of ``contour pinches'', which is explained in the appendix. There we also show, that there is no pole canceling the zero $\G(\de_2)^{-1}$ and we conclude that there are no contributions from the non-evaluated integrals to the $\al$ four-point function at $d=3$.

Therefore the only contributions to the $\al$ four-point function arise from the coefficients $c_{nm}^{(2j)},\; j=1,2,$ which have been extracted for general $d$ in \cite{Lang:1992pp}. Applying this result for $d=3$ gives for the connected part 
\begin{multline}\label{prelimeq}
\langle \al(x_1) \cdots \al(x_4) \rangle \Bigl \rvert_{\textrm{conn.}, d=3} = (x_{12}^2 x_{34}^2)^{-2} \frac{1}{N} \Bigl[-u^{-\frac{3}{2}}(1+u-v) + (uv)^{-\frac{3}{2}}(1-u-v) \Bigr].
\end{multline}
However, this expression is not crossing symmetric, i.e. invariant under the replacements
\begin{align}
(a) & \qquad 1 \leftrightarrow 2; \qquad u \leftrightarrow v \\
(b) & \qquad 2 \leftrightarrow 3; \qquad u \mapsto \frac{1}{u},\; v \mapsto \frac{v}{u}.
\end{align}
In fact, (\ref{prelimeq}) contains only contributions of the graphs $B_{21}$ and $B_{22}$ but not of $B_{23}$, because they were found to vanish in the $(u,1-v)$ expansion. Nevertheless we find from crossing symmetry the complete expression
\begin{align}\label{complete}
(x_{12}^2 x_{34}^2)^{-2} \frac{1}{N} \Bigl[-u^{-\frac{3}{2}}(1+u-v) - v^{-\frac{3}{2}} (1-u+v) + (uv)^{-\frac{3}{2}}(1-u-v) \Bigr].
\end{align}
Thus we conclude that the analytic continuation from the $(u,1-v)$ to the $(v,1-u)$ expansion produces (by a Kummer relation for a $_2F_1$ series) a pole at $d=3$ canceling the zero.

Let us introduce the following notation for the composite fields: We write
\begin{align}
(n \phi)_{l,t}, \; (m \al)_{l,t}
\end{align}
for the composite field of $n$ $\phi$-fields and $m$ $\al$-fields, respectively, of spin $l$ and twist $t$. Then the  result (\ref{complete}) can be expanded in the $(1,3)-(2,4)$ channel (``s-channel'') into exchange amplitudes of currents $(2 \phi)_{l,t=0}$ from the first and third term and exchange amplitudes of composites $(2 \al)_{l,t}$ from the second term 
\begin{multline}
\langle \al(x_1) \cdots \al(x_4) \rangle \Bigl \rvert_{\textrm{conn.}, d=3} \\
= \sum_{l \, \textrm{even}} \gamma_l^2 \qquad \;
\begin{picture}(40,30)
\put(0,-17){\dashline{4}(-24,1)(0,20)
\dashline{4}(-24,39)(0,20)
\put(0,21){\line(1,0){40}}
\put(0,19){\line(1,0){40}}
\put(2,8){$(2 \phi)_{l,t=0}$}
\dashline{4}(40,20)(63,0)
\dashline{4}(40,20)(63,39)}
\end{picture}
\; \qquad + \sum_{\substack{l \, \textrm{even} \\ t}} \kappa_{l,t}^2\;\qquad 
\begin{picture}(40,30)
\put(0,-17){\dashline{4}(-24,1)(0,20)
\dashline{4}(-24,39)(0,20)
\put(5,8){$(2\al)_{l,t}$}
\put(0,1){\dashline{4}(0,20)(40,20)}
\put(0,-1){\dashline{4}(0,20)(40,20)}
\dashline{4}(40,20)(63,0)
\dashline{4}(40,20)(63,39)}
\end{picture} 
\qquad \quad .
\end{multline}
The latter do not contain any logarithmic terms ($\log u$) and correct only the normalization of the corresponding exchange amplitudes from the disconnected graphs $A_1, A_2, A_3$. 

We conclude this section with a short discussion on arbitrary $n$-point functions of $\al$-fields till order $O(1/N)$. For $n>4$ we get only disconnected contributions to this order. If $n$ is even, at order $O(1/N)$ we have $\frac{n-4}{2}$ $\al$-propagators and one four-point insertion, which is taken from (\ref{complete}). If $n$ is odd, we have a three-point insertion, whose contribution at $O(1/N^{\frac{1}{2}})$ vanishes. Therefore there are no contributions of $n$-point functions with odd $n$ at order $O(1/N)$.


\section{Anomalous dimensions at $\boldsymbol{d=3}$}

Each of the exchanged currents $J_l$, $l=4,6,8,\ldots$ acquires an anomalous dimension, which is of order $O(1/N)$, but contributes to the $\al$ four-point function at order $O(1/N^2)$. It can be calculated from the four-point function of the $\phi$-field and gives
\begin{multline}
\frac{\eta(J_l)}{\eta(\phi)} = \frac{2(l+2)}{(2 l-1)(2l+1)} \Bigl[2(l-1) +\sum_{p=1}^{\frac{1}{2} l-2} \bigl((p+1)!\bigr)^2 \binom{l}{p+1} \frac{(4+p)_{l-4-2p}}{(4)_{l-4}} \Bigr] + O(1/N^2),
\end{multline}
see eq. (3.36) in \cite{LR}. 

The absence of $\log u$ terms in any $n$-point function of $\al$-fields till order $O(1/N)$ shows that composite fields of arbitrary many $\al$-fields are constructed like composites of quasi-free fields. This means that such a composite field $(n \al)_{l,t}$ is formed by $l+2t$ differentiations and results in a tensor field of rank $l$. Its dimension is given by
\begin{align}\label{anomdimofcomp}
\de((n \al)_{l,t}) = n \bigl( 2+\eta(\al) \bigr) + l +2t +O(1/N^2).
\end{align}
There are many references in the literature for anomalous dimensions of such composites $(n \al)_{l,t}$ for general $2<d<4$, from which we can check (\ref{anomdimofcomp}) at $d=3$. 

The unique scalar field $(n \al)_{0,0}$ has anomalous dimension
\begin{multline}
\de\bigl( (n \al)_{0,0} \bigr) = 2 n -\frac{4}{N} \eta_1(\phi) \frac{2\mu-1}{\mu-2} \Bigl[n(n-2)(\mu-1) \\ + \binom{n}{2} (2 \mu-1) (\mu-2) \Bigr] + O(1/N^2),
\end{multline}
see eq. (3.33) in \cite{Lang:1993ct}. It is remarkable that for $d=3$ the quadratic dependence of the anomalous dimension on $n$ reduces to a linear one, and that we get the special case $l=t=0$ of (\ref{anomdimofcomp}). 

For $l\not= 0, t=0$ and arbitrary $n$ we must expect that several different conformal fields with the same quantum numbers $l,t,n$ have different anomalous dimensions. We can use eqs. (3.28)-(3.42) and (3.17)  of \cite{Lang:1993ct} to check the case $d=3$. We find indeed (\ref{anomdimofcomp}) with the degeneracy lifted presumably at $O(1/N^2)$. For quasi-free fields there is a cohomology approach to the construction of different conformal fields \cite{Lang:1994tu} which can be applied to this case, too.


\section*{Appendix}

Consider the example 
\begin{multline}\label{example}
\frac{1}{2\pi i} \int_{-i\infty}^{i \infty} \ind s \, \G(\al+s) \G(\beta +s) \G(\gamma - s) \G(\de -s) \\ = \frac{ \G( \al + \gamma) \G( \beta+\gamma) \G(\al+\de) \G(\beta+\de)}{\G(\al+\beta+\gamma+\de)}
\end{multline}
The integrand has two sequences of poles
\begin{align}\label{rightsequence}
\gamma-s & = -N_1 \nonumber \\
\de-s &= -N_2, \quad(N_{1,2} \in \mathbb{N}_0)
\end{align}
tending to $+\infty$, and two sequences of poles
\begin{align}\label{leftsequence}
\al +s &= -M_1 \nonumber \\
\beta+s &= -M_2 \quad (M_{1,2} \in \mathbb{N}_0)
\end{align}
tending to $- \infty$ in the $s-$plane. The integral contour is chosen in such a way that all poles from the ``left sequence'' (\ref{leftsequence}) are to the left of the contour, and all poles of the `` right sequence'' (\ref{rightsequence}) are to its right. In the case that such a choice is not possible, namely if a left and a right series intersect in a (necessarily finite) number of points, the contour is said to be pinched in these points and poles arise there. This pinching occurs in the following cases
\begin{align}
s &= \gamma + N_1 = -\al-M_1, &\textrm{i.e. if}\; &\al+\gamma = -N_1-M_1 \\
s &= \gamma + N_1 = -\beta-M_2, & \textrm{i.e. if}\;  &\beta+\gamma = -N_1-M_2 \\
s &= \de + N_2 = -\al-M_1, &\textrm{i.e. if}\; & \al+\de = -N_2-M_1 \\
s &= \de + N_2 = -\beta-M_2, &\textrm{i.e. if}\; & \beta+\de = -N_2-M_2 
\end{align}
Thus we conclude that the possible poles are just those of the numerator of the result in (\ref{example}), i.e. the poles of
\begin{align}
\G ( \al + \gamma ) \G ( \beta + \gamma ) \G ( \al + \de ) \G ( \beta + \de ), 
\end{align}
and that the integrand is equal to this meromorphic function times an arbitrary holomorphic function, as (\ref{example}) shows.

Now we treat $c_{nm,1}^{(2)}$ from eq. (\ref{crucint}) in this fashion. The relevant integral is 
\begin{multline}
\frac{1}{(2 \pi i)^2} \int \ind x \int \ind y \,\G \Bigl[ {-x, -y , \al_4+n+m+y, \beta_3-\al_2-n+x,  \beta_4-\al_4-y \atop \mu -\al_4-y, \mu+\al_2- \beta_3+2n+m-x} \Bigr] \\ 
\G(  \al_3-\beta_1-x, \mu-\beta_2+x+y, \mu-\beta_1+n-x-y ) (\mu-\beta_3-x)_n 
\end{multline}
Note that the $\G-$ function has no zeros. With 
\begin{align}
a &= \mu-\beta_1+n \nonumber \\
b &= \beta_3-\al_2-n \nonumber \\
c &= \mu -\beta_2 \nonumber \\
d &= \al_3-\beta_1
\end{align}
we get the right sequences of poles
\begin{align}
- x &= -M_1 \nonumber \\
a-x-y &= -M_2 \nonumber \\
d-x &= -M_3, \quad (M_{1,2,3} \in \mathbb{N}_0),
\end{align}
and the left sequences
\begin{align}
b+x &= -N_1 \nonumber \\
c+x+y &= -N_2, \quad (N_{1,2} \in \mathbb{N}_0),
\end{align}
which combine into six pinch sequences of $x$
\begin{align}
b &= -N_1-M_1 \nonumber \\
c+y &= -N_2-M_1 \nonumber \\
a+b-y &= -N_1-M_2 \nonumber \\
a+c &= -N_2-M_2 \nonumber \\
b+d &= -N_1-M_3 \nonumber \\
c+d+y &= -N_2-M_3,
\end{align}
leading to 
\begin{align}
\G(b,a+c,b+d,c+y,c+d+y,a+b-y).
\end{align}
Next we consider the $y-$pinches of 
\begin{align}
\G( -y, e+y, f-y, c+y, c+d+y, a+b-y),
\end{align}
where
\begin{align}
e &= \al_4+n+m \nonumber \\
f &= \beta_4 -\al_4
\end{align}
From this we read off the right sequences of poles
\begin{align}
-y &= -R_1 \nonumber \\
f-y &= -R_2 \nonumber \\
a+b-y &= -R_3 ,\quad (R_{1,2,3} \in \mathbb{N}_0)
\end{align}
and the left sequences
\begin{align}
e+y &= -S_1 \nonumber \\
c+y &= -S_2 \nonumber \\
c+d+y &= -S_3, \quad (S_{1,2,3} \in \mathbb{N}_0).
\end{align}
Then we obtain pinches in the following cases
\begin{align}
e &= -S_1-R_1  \nonumber \\
e+f &= -S_1-R_2 \nonumber \\
e+a+b &= -S_1-R_3 \nonumber \\
c &= -S_2-R_1 \nonumber \\
c+f &= -S_2-R_2 \nonumber \\
c+a+b &= -S_2-R_3 \nonumber \\
c+d &= -S_3-R_1 \nonumber \\
c+d+f &= -S_3-R_2 \nonumber \\
c+d+a+b &= -S_3-R_3,
\end{align}
with the corresponding meromorphic factor
\begin{multline}
\G( b, c, e, a+c, b+d, c+d, c+f, e+f,c+d+f,c+a+b)\\
\G(e+a+b,a+b+c+d).
\end{multline}
This does not include a pole at $\mu =3/2$, which would be necessary to cancel the simple zero in the prefactor $\G(\de_2)^{-1}$ in (\ref{crucint}).

\end{document}